# Low temperature dielectric anomalies in HoMnO$_3$:
# The complex phase diagram

F. Yen[1], C. R. dela Cruz[1], B. Lorenz[1], Y. Y. Sun[1], Y. Q. Wang[1], M. M. Gospodinov[2], and C. W. Chu[1,3,4]

[1] *TCSUH and Department of Physics, University of Houston, Houston, Texas 77204-5002, USA*

[2] *Institute of Solid State Physics, Bulgarian Academy of Sciences, 1784 Sofia, Bulgaria*

[3] *Lawrence Berkeley National Laboratory, 1 Cyclotron Road, Berkeley, California 94720, USA*

[4] *Hong Kong University of Science and Technology, Hong Kong, China*

## Abstract

The dielectric constant of multiferroic hexagonal HoMnO$_3$ exhibits an unprecedented diversity of anomalies at low temperatures (1.8 K < T < 10 K) and under external magnetic fields related to magnetic phase transitions in the coupled system of Ho moments, Mn spins, and ferroelectric polarization. The derived phase diagram is far more complex than previously assumed including reentrant phases, phase transitions with distinct thermal and field hysteresis, as well as several multicritical points. Magnetoelastic interactions introduce lattice anomalies at the magnetic phase transitions. The re-evaluation of the T-H phase diagram of HoMnO$_3$ is demanded.

Among the multiferroic compounds hexagonal rare-earth manganites, $R$MnO$_3$ ($R$=Sc, Y, Ho to Lu), have attracted recent attention because of the coexistence of antiferromagnetic (AFM) and ferroelectric (FE) orders below their Néel temperatures, $T_N$ <100 K. The mutual interference and the possible correlation of magnetic and ferroelectric orders are of fundamental interest and of significance for prospective applications. Since the first report[1] of a dielectric anomaly at $T_N$ of YMnO$_3$ and later observations of similar effects in other hexagonal $R$MnO$_3$ the origin of the magneto-dielectric coupling in these compounds has been a matter of intense discussions.[2–5] In the magnetic symmetries realized just below $T_N$ the direct coupling between the in-plane staggered AFM magnetization and the c-axis FE polarization is not allowed.[6] The indirect coupling via lattice strain or other secondary effects has there-fore been proposed.[5] In fact, the existence of strong spin-phonon interactions was concluded from the anomalous temperature dependence of the thermal conductivity[7] and the thermal expansion anomalies observed at $T_N$ of HoMnO$_3$.[8]

The magnetic order below 100 K is very complex for some $R$MnO$_3$ and it is determined by the strong in-plane AFM superexchange interaction of the Mn$^{3+}$ spins and their geometric frustration in the *P*6$_3$*cm* hexagonal structure. While strong in-plane magnetic correlations exist well above the Néel temperature[8,9] the long range AFM order of the Mn$^{3+}$ sets in only below 100 K with a frustrated, non collinear spin arrangement (neighboring moments form an angle of 120º). The different spin arrangements and magnetic symmetries compatible with the *P*6$_3$*cm* crystalline structure have been discussed in detail by Fiebig et al.[10] The direct magnetic exchange of the $R^{3+}$ ions, essential at low temperatures, and the coupling of the rare earth moments with the Mn spins as well as the FE polarization gives rise to a wealth of interesting physical phenomena and a complex magnetic phase diagram. Several subsequent magnetic phase transitions have been observed in HoMnO$_3$ below $T_N$ =76 K, including a sharp Mn spin rotation transition at $T_{SR}$=33 K with the onset of AFM Ho-moment order and another transition at about 5 K involving a major increase of the Ho$^{3+}$ moments.[4,10–13] Different magnetic symmetries partially coexisting below $T_N$ have also been reported for LuMnO$_3$ and ScMnO$_3$.[13]

Under applied magnetic fields HoMnO$_3$ exhibits the most complex T-H phase diagram of all hexagonal $R$MnO$_3$ with several field-induced re-entrant phases and transitions at low temperatures.[5,10,14] Distinct anomalies of the dielectric constant ε have been observed at the zero field transitions, notably a sharp peak of ε(T) at $T_{SR}$ that defines a new reentrant phase induced by magnetic fields.[5] This phase was shown to extend over a broader range in the T-H phase diagram.[5,14] Whereas the previous dielectric investigation was limited to temperatures above 7 K measurements of magnetization and heat capacity have revealed additional phase boundaries in the (low T)-(low H) region of the phase diagram.

In order to unambiguously identify the various phase boundaries and to resolve the complex phase diagram of HoMnO$_3$ we have extended the investigation of the dielectric proper-ties to low temperatures (T>1.8 K) and high magnetic fields (H<70 kOe). The dielectric constant as a function of temperature and magnetic field exhibits sharp anomalies at all magnetic phase transitions due to the strong magneto-dielectric coupling in the compound. New phase transitions are detected at low temperatures. Upon the completion of the present work, we learned that a neutron scattering study has been carried out by Vajk et al.[15] showing also a complex T-H phase diagram with some distinctly different phase boundaries from ours. The disagreement shows that more explorations of the phase diagram and the magnetic structures by alternative experimental methods are warranted.

Single crystals of HoMnO$_3$ of typical size 3 x 5 x 0.1 $mm^3$ were synthesized as described elsewhere.[14] The capacitance was measured with the electric field directed along the hexagonal *c*-axis employing the high precision capacitance bridge AH 2500A (Andeen Hagerling) operating at 1 kHz frequency. The Physical Property Measurement System (Quantum Design) was used to control the temperature and magnetic field (parallel to the *c*-axis).

The dielectric constant below 100 K at zero magnetic field exhibits three distinct anomalies associated with the known magnetic transitions of HoMnO$_3$ at 76, 33, and 5.2 K, respeticely.[5,10,12,14,16] The low temperature anomaly of ε(T) at $T_2$=5.2 K and its magnetic field dependence is shown in Fig. 1. The sharp rise at $T_2$ is followed by a continuous decrease of ε(T) to lower T . The small hump at 4.9 K (denoted by $\tilde{T}_1$ and the dotted line in Fig. 1a) indicates another subtle change in the magnetic structure. A second small peak of the specific heat was observed at the same temperature.[14] Therefore, the low temperature magnetic transition of HoMnO$_3$ appears to proceed in two subsequent steps, only 0.3 K apart. Neutron scattering experiments suggest a major increase of the Ho magnetic moments at this temperature while the Mn spins rotate another time by 90º.[12] These scattering measurements, however, do not have the temperature resolution to separate the two transitions that are unambiguously shown in the data of Fig. 1a. The two anomalies of ε(T) separate further apart in an external magnetic field and $\tilde{T}_1$ which is pushed to lower temperatures (dotted line in Fig. 1a) is unambiguously determined up to about 14 kOe by the change of slope of ε(T).

Above 5 kOe a third anomaly of ε(T) arises at a lower $T_3$ in the form of a sharp increase, as shown in Fig. 1a. $T_3(H)$ increases with H and the anomaly grows and peaks about 3.3 K and 12 kOe. All data have been acquired upon decreasing and increasing temperature. No thermal hysteresis was observed at the transitions ($T_2$, $\tilde{T}_1$, and $T_3$) for H < 10 kOe. At 11 kOe, however, a narrow hysteresis is evidenced near 6 K (Fig. 1a). With further increasing H a small step-like anomaly of ε(T) appears at $T_5$ below $T_2$ where a drop of ε occurs. Such a small anomaly shifts to lower T with increasing field (Fig. 1b). $T_5(H)$ is clearly associated with a thermal hysteresis and the ε-anomaly broadens with increasing H. Above 15 kOe the exact value of $T_5$ is difficult to extract from ε(T) because of the large hysteresis. Between 15 and 18 kOe the transition is only observed upon heating whereas with cooling $T_5$ merges with $T_3$ and enters the low temperature phase ("LT1")[14] at $T_3 \approx 3.2$ K, as indicated by the arrows attached to the 16 kOe data in Fig. 1b. The exact values of $T_5(H)$ and the width of the hysteresis associated with $T_5$ was more accurately derived from isothermal ε(H) scans.[17] Above 18 kOe a distinct peak of ε(T) arises at about 2.8 K (18.7 kOe). The position of this peak, shown in Figs. 1b and 1c (denoted by $\tilde{T}_3(H)$), moves to higher T with increasing field. The peak of ε(T) broadens with increasing H and it becomes less well resolved at fields exceeding 28 kOe (Fig. 1c). At 22 kOe two additional steps of ε(T) appear below 2.5 K indicating other sharp transitions in the phase diagram of HoMnO$_3$. One of these transitions coincides with the $T_1(H)$ phase boundary separating the "INT" and "HT1" phases. The other sudden change of ε(T), denoted by $T_4(H)$ in Figs. 1c and 1d, can be followed to the highest magnetic fields of 70 kOe and it is obviously related to the transition into the "LT2" phase (the AFM-A$_1$ phase in the notation of Fiebig et al.).[10,14] It is remarkable that this transition is very sharp and does not show indications of thermal hysteresis in contrast to conclusions from optical experiments.[10] $T_4$ first increases from H=22 kOe, passes through a maximum at 52 kOe and 4.6 K, and decreases again at higher magnetic field (Figs. 1c and 1d). $T_1(H)$ and $T_4(H)$ seem to merge at low temperatures close to 20 kOe. Some of these phase boundaries are consistent with those determined magnetically, calorimetrically, and optically.[10,14]

The isothermal $\varepsilon(H)$ displays similar anomalies as $\varepsilon(T)$ and allows the determination of the phase boundary $T_5(H)$ and the hysteretic regions in the phase diagram more precisely. The data are included as open symbols in Fig. 2. In addition to the anomalies of $\varepsilon(T)$ discussed above a broader maximum of $\varepsilon(H)$ appears below 12 kOe between 3.2 and 4.8 K. This maximum of $\varepsilon(H)$, shown as the dashed line $\tilde{T}_2$ in Fig. 2, may not represent another phase boundary but it indicates a smooth change of the magnetic structure in this region of the phase diagram. Details will be published elsewhere.[17]

The dielectric anomalies observed in the T- and H-dependence of $\varepsilon$ are closely correlated with similar anomalies of the ac magnetic susceptibility $\chi_{ac}$.[14] The similarities include the enhancement of both susceptibilities in the "INT" phase, the remarkable peak at $\tilde{T}_3$, the broad peak along $\tilde{T}_2(H)$, and the sharp changes in crossing the $T_1(H)$, $T_2(H)$, and $T_3(H)$ phase boundaries. The common features of $\varepsilon$ and $\chi_{ac}$ prove the existence and the strength of the magneto-dielectric coupling in HoMnO$_3$ at low T. Due to this coupling the magnetic phase diagram is completely determined by the anomalies of $\varepsilon$ as shown above and summarized in Fig. 2.

The multitude of anomalies observed in HoMnO$_3$ by magnetic, calorimetric, dielectric as well as optical and neutron scattering experiments makes it difficult to unambiguously identify the various phases and transition lines. A systematic approach appears necessary to reconcile all data correctly. Based on the sharp, step-like changes of $\varepsilon$ we have identified five phase boundaries denoted by $T_1(H)$ to $T_5(H)$ (solid lines with symbols in Fig. 2). Other remarkable features of $\varepsilon(T,H)$ such as narrow or broader peaks and the change of slope of $\varepsilon(T)$ denoted by $\tilde{T}_1$ to $\tilde{T}_3$ are shown as dashed lines in Fig. 2. Currently it is not clear if the latter anomalies define true phase boundaries but they indicate possible regions of crossover associated with smooth changes of the magnetic structure. In order to resolve the different phase boundaries in the very complex (low T)-(low H) region we approach this part of the phase diagram from high temperature or high magnetic field where the magnetic phases are well defined.

Based on the dielectric[5], magnetic and thermodynamic[14] anomalies $T_1(H)$ and $T_2(H)$ define the transitions between the "HT1" and "HT2" phases including the intermediate "INT" phase. $T_1$ is clearly defined by the sharp drop of $\varepsilon$ and $\chi$ and it can be traced from 33 K (H=0) to about 2.5 K (22 kOe) where it merges with $T_4$. $T_2(H)$ is uniquely determined by the sharp increase of $\varepsilon$, $\chi$, and the heat capacity[14] and it stretches from (33 K, H=0) to (20 K, 34 kOe) and back to (5.2 K, H=0), as shown in Fig. 2. The dome shaped "LT1" phase and its phase boundary $T_3(H)$ are unambiguously defined by the step of $\varepsilon(T,H)$ as well as by the narrow peaks of $\chi(T,H)$ and the heat capacity.[14] $T_5(H)$ connecting the $T_1$ and $T_3$ phase boundaries near 15 kOe is clearly identified by the small step of $\varepsilon(T)$ and its associated hysteresis. Similar features have been observed in $\chi_{ac}$.[17] The remaining phase boundary between the "HT1" and "LT2" phases, $T_4(H)$, was not well characterized before because of hysteresis effects observed in optical experiments[10] or due to broad changes of the magnetization.[14] The dielectric data, however, prove that this phase transition is sharp and no hysteresis was observed. $T_4$ defines another dome shaped phase with a maximum of 4.6 K at 52 kOe. The weaker anomalies of $\varepsilon(T,H)$ labeled $\tilde{T}_1$ to $\tilde{T}_3$ (dashed lines in Fig. 2) will be discussed elsewhere.[17]

The phase boundaries $T_1$ to $T_5$ include several multicritical points. Two tricritical points are located at the intersection of $T_5(H)$ with $T_2(H)$ and $T_3(H)$. Of particular interest is the region near 20 kOe and below 3 K. $T_1$ and $T_4$ come close and could form another tricritical point. However, $T_3(H)$ is also very close and there is the possibility that all three transition lines merge at lower T, possibly at T=0. This would result in a tetracritical point at or close to zero temperature, an unusual phenomenon of considerable interest. Investigations at ultra low temperatures are therefore desired to elucidate this remarkable T-H region of the phase diagram.

The phase diagram summarized in Fig. 2 is to be compared with the recent conclusions from neutron scattering data.[15] Although most of the anomalies observed in the scattering intensities can be accommodated by our phase diagram the major phase boundaries are drawn differently resulting in an unusual pentacritical point at about 6 K and 13 kOe.[15] The main difference to our results is a hysteretic anomaly between 10 and 15 kOe derived from the (2,1,0) and (1,0,0) peak intensities that extends from about 5 K to lowest T. This anomaly appears to be the extension of the $T_5(H)$ phase boundary into the "LT1" phase, however, we have not observed any anomaly of $\varepsilon(H)$ or $\chi_{ac}(H)$ below $T_3(H)$ between 5 and 20 kOe. The possible origin of this discrepancy could lie in the different sample orientations examined by the two experiments. $\varepsilon$ and $\chi$ have been measured with electric and magnetic fields along the c-axis whereas the neutron scattering experiments do not include out-of-plane magnetic peaks.

The physical nature and the magnetic structures of the different phases revealed in Fig. 2 are not completely resolved. Neutron scattering[12,16] and optical investigations[10] have assigned the $P6_3cm$, $P\underline{6}_3c\underline{m}$, and $P6_3cm$ magnetic symmetries to "HT1", "HT2", and the phase below 5 K (enclosed by $T_2$, $T_5$, and $T_3$ in Fig. 2), respectively. For the "INT" phase we propose the lower $P\underline{6}_3$ symmetry where the Mn-spins form an angle $\varphi_{Mn}$ between 0 and 90° with the *a*-axis. We suggest that $\varphi_{Mn}$ changes continuously as a function of T and H within the "INT" phase. The magnetic structures of the "LT1" and "LT2" phases are yet to be resolved.

The intriguing simultaneous occurrence of the magnetic and dielectric anomalies raises the question about the nature of their correlation. Strong magnetoelastic effects have recently been shown to exist in $HoMnO_3$ at H=0 and to be responsible for the magneto-dielectric coupling at and below $T_N$.[8] To investigate lattice anomalies in the low temperature region of the phase diagram we have measured the *c*-axis length as function of H for constant T between 1.8 and 20 K (Fig. 3) using the strain gage method. At higher T's (Fig. 3a) $c(H)$ shows a rapid increase with H between $T_2$ and $T_1$ in the "INT" phase. The corresponding slope changes of $c(H)$ coincide with the $T_2$ and $T_1$ phase boundaries of Fig. 2. The increase of the *c*-axis with the transition from "HT2" to "HT1" is consistent with the zero field expansion measurements that showed an abrupt increase of the *c*-axis length with increasing T at $T_{SR}$.[8] Below 5 K $c(H)$ is a more complex function (Fig. 3b). The slope change at $T_1(H)$ becomes more pronounced and, below about 3 K, two sharp minima of $c(H)$ develop between 5 and 20 kOe. These two minima precisely trace the phase boundary $T_3(H)$ of Fig. 2. No other anomalies could be observed within the experimental resolution. $c(H)$ is constant in the "LT2" phase (H>22 kOe, T<3 K), i.e. there is no *c*-axis magnetoelastic effect in this phase. The observed lattice anomalies prove the important role of the spin lattice coupling at low temperatures in HoMnO$_3$ and they provide a possible explanation for the observed correlation of magnetic and dielectric properties and the complex T-H dependence of the dielectric constant.

In summary the T-H phase diagram of HoMnO$_3$ is resolved via measurements of the dielectric constant. At least 6 different phases can be distinguished below T$_N$ through sharp changes of ε(T,H). Additional anomalies in form of a slope change of ε(T) or distinct peaks of ε(T,H) are monitored throughout the phase diagram. Several multicritical points are described and the possible existence of an unusual tetracritical point at zero temperature is proposed. The strong correlation of magnetic and dielectric properties with the lattice anomalies observed at the magnetic phase boundaries emphasize the role of magnetoelastic effects in promoting the magneto-dielectric coupling in HoMnO$_3$. The exact nature of the many magnetic phases in the T-H diagram has yet to be resolved.

**Acknowledgments**

This work is supported in part by NSF Grant No. DMR-9804325, the T.L.L. Temple Foundation, the J. J. and R. Moores Endowment, and the State of Texas through the TCSUH and at LBNL by the DoE. The work of M. M. G. is supported by the Bulgarian Science Fund, grant No F 1207.

---

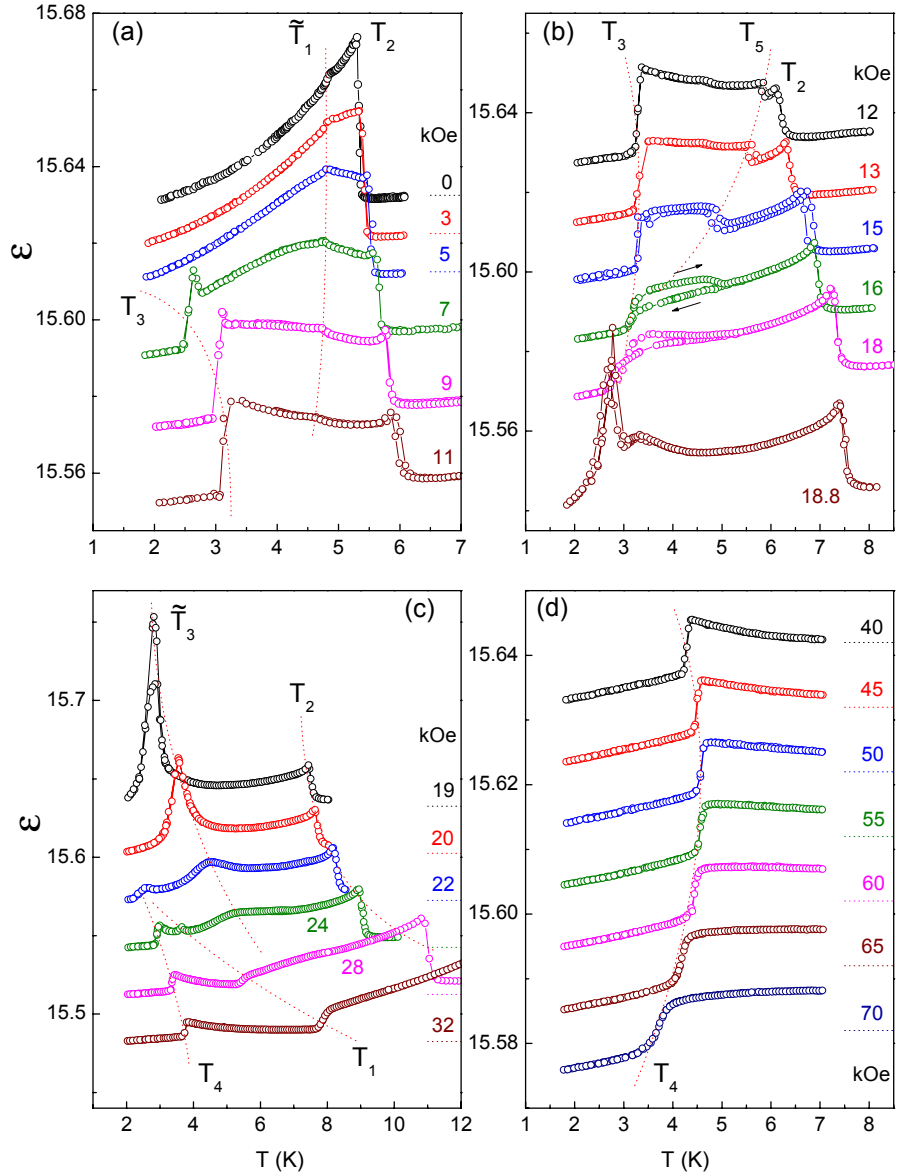

**FIG. 1**: (Color online) Low temperature dielectric constant, ε(T), at different magnetic fields. For better clarity the curves are vertically offset, the base line for each curve is shown by the dotted line with the magnetic field value as a label. The phase boundaries and anomalies discussed in the text are indicated by dotted lines labeled as $T_i$ and $\tilde{T}_j$.

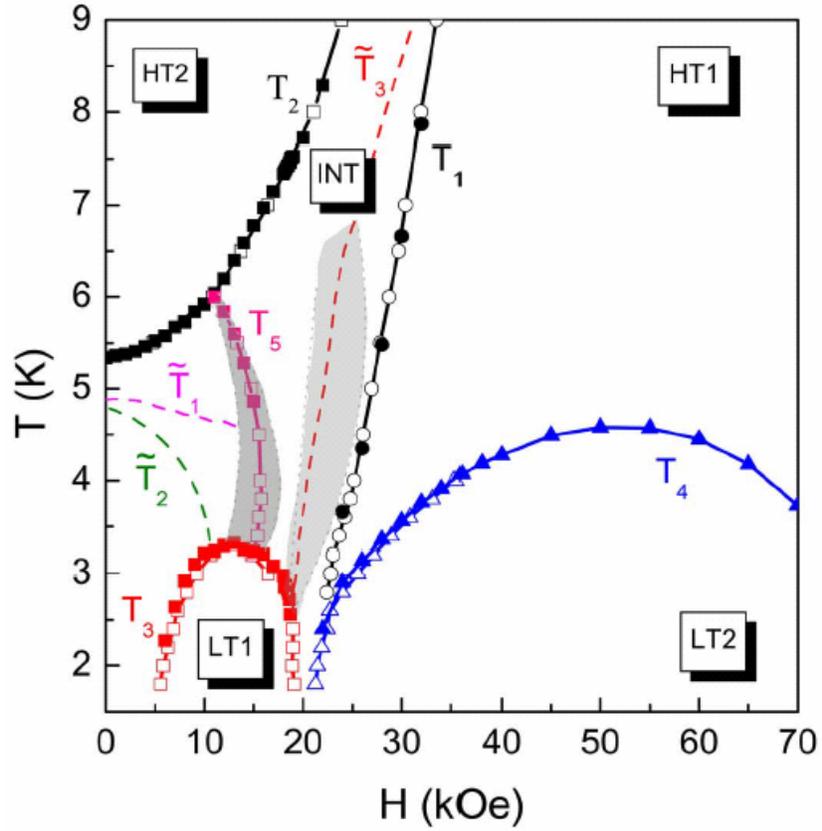

**FIG. 2**: (Color online) Low temperature phase diagram of HoMnO$_3$. Phase boundaries $T_1$ to $T_5$ are drawn as solid lines. Closed and open symbols are derived from $\varepsilon(T)$ and $\varepsilon(H)$, respectively. Additional anomalies $\tilde{T}_1$ to $\tilde{T}_3$ are indicated by dashed lines. The shaded areas mark regions in the phase diagram where hysteresis effects have been observed.

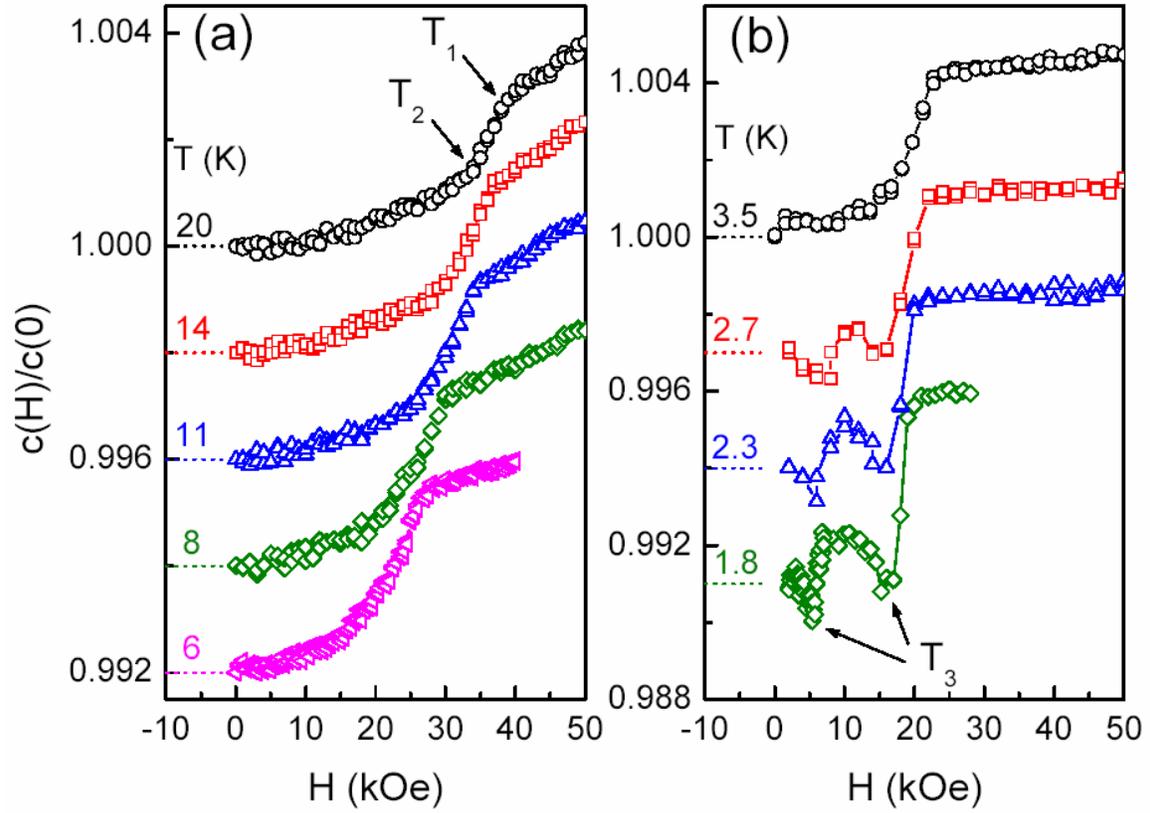

**FIG. 3**: (Color online) Magnetic field dependence of the *c*-axis length at various temperatures. Clear anomalies are observed at the transition temperatures $T_1$, $T_2$, and $T_3$. Different data sets are offset for clarity. The dotted line with the temperature value attached marks the H=0 value for each curve.